\documentclass[a4paper,11pt]{article}

\usepackage{placeins}
\usepackage{fourier}
\usepackage{color}
\usepackage{hyperref}
\usepackage{booktabs}
 \usepackage{graphicx}
\usepackage{url}
\usepackage[affil-it]{authblk}
\usepackage{amsmath}
\usepackage{wrapfig}
\usepackage{multirow}
\usepackage[T1]{fontenc}
\usepackage{times}
\usepackage[caption=true]{subfig}

\begin{document}

\title{Assessing Intra-class Diversity and Quality of Synthetically Generated Images in a Biomedical and Non-biomedical Setting\thanks{This work is supported by the Munster Technological University's Risam Scholarship Award}}

\author{Muhammad Muneeb Saad, Mubashir Husain Rehmani, and Ruairi O'Reilly}
\affil{Munster Technological University, Cork, Ireland.}
\date{}
\maketitle
\thispagestyle{empty}

\begin{abstract}
In biomedical image analysis, data imbalance is common across several imaging modalities. Data augmentation is one of the key solutions in addressing this limitation. Generative Adversarial Networks (GANs) are increasingly being relied upon for data augmentation tasks. Biomedical image features are sensitive to evaluating the efficacy of synthetic images. These features can have a significant impact on metric scores when evaluating synthetic images across different biomedical imaging modalities. Synthetically generated images can be evaluated by comparing the diversity and quality of real images. Multi-scale Structural Similarity Index Measure and Cosine Distance are used to evaluate intra-class diversity, while Frechet Inception Distance is used to evaluate the quality of synthetic images. Assessing these metrics for biomedical and non-biomedical imaging is important to investigate an informed strategy in evaluating the diversity and quality of synthetic images. In this work, an empirical assessment of these metrics is conducted for the Deep Convolutional GAN in a biomedical and non-biomedical setting. The diversity and quality of synthetic images are evaluated using different sample sizes. This research intends to investigate the variance in diversity and quality across biomedical and non-biomedical imaging modalities. Results demonstrate that the metrics scores for diversity and quality vary significantly across biomedical-to-biomedical and biomedical-to-non-biomedical imaging modalities.  
\end{abstract}

\section{Introduction}

In image analysis, data imbalance affects datasets with asymmetrical distribution of samples for classes within a dataset. In the domain of biomedical imaging, this data imbalance is particularly common for classes of rarer diseases and conditions \cite{qin2022learning}. When a model is exposed to an imbalanced distribution during training, it may focus on the majority class and fail to learn important features and patterns from the minority class. Consequently, when the model encounters new, unseen data, it may struggle to classify samples from the underrepresented classes accurately. To address this issue, data augmentation is utilized to generate new samples from existing data. The intent is to improve the model's ability to generalize and therefore, accuracy, thus contributing to better clinical outcomes \cite{zhao2021improved}. 

Generative Adversarial Networks (GANs) are particularly useful for data augmentation because they can generate synthetic images that contain a similar distribution of diversified features, such as shapes, textures, or colors, compared to real images \cite{allahyani2023divgan}. For deep learning models, GANs can increase the diversity of the training dataset and improve classifiers' performance by generating these synthetic images. GANs are popular in the biomedical imaging domain due to their ability to synthesize realistic images from a given distribution \cite{zhao2021improved}. GANs consist of two models based on neural networks: a generator and a discriminator, which work together in a game-theoretic setting to produce realistic images. In biomedical image analysis, diverse image features play a significant role in training a model to produce better clinical diagnostic outcomes \cite{segal2021evaluating}. The architecture of a GAN is designed such that it can learn these diversified features and synthesize them in synthetic images.

The diversity and quality of synthetic images are important considerations throughout the training of GANs. In the context of GANs, diversity refers to the variation and range of generated image samples produced by a GAN model. It indicates a GAN's ability to generate a diverse set of synthetic images that cover different aspects, styles, and variations present in the original training images \cite{allahyani2023divgan}. Intra-class diversity refers to the diversity assessment within a single class of images. Similarly, the quality of synthetic images refers to the fidelity, realism, and perceptual similarity of the generated image samples compared to the real data. It indicates a GAN's ability to produce high-quality synthetic image data that closely resembles the distribution of the original training data \cite{he2020retinal}. 

The intra-class diversity of synthetic images is evaluated using the Multi-scale Structural Similarity Index Measure (MS-SSIM) \cite{odena2017conditional} and Cosine Distance (CD) \cite{borji2019pros}. The quality of synthetic images is evaluated using the Frechet Inception Distance (FID) \cite{borji2019pros}. 

It is critical to assess the intra-class diversity and quality of synthetically generated biomedical images. Biomedical images contain complex and diverse features indicating vital information about the subject. Evaluating these synthetic images should consider examining the diverse and high-quality features across different biomedical imaging modalities.

MS-SSIM, CD, and FID are significantly dependent on the salient features inherent in an imaging modality. These metrics use image features such as textures, luminance, and orientation of objects for quantification of the diversity and quality of synthetic images. Consequently, score values for these metrics vary across different biomedical imaging modalities and non-biomedical images. Furthermore, quantifying the diversity and quality of synthetic images can be impacted by the number of image samples. Therefore, the assessment of MS-SSIM, CD, and FID metrics is important for biomedical and non-biomedical imaging domains. It enables the efficacy of evaluating GAN architectures in generating diversified and high-quality synthetic images. 

The contribution of this work is as follows:

\begin{itemize}
    \item Study the effect of sample size on evaluation metrics used for assessing the diversity and quality of synthetic images, in both non-biomedical and biomedical imaging fields.
    \item Examine the inconsistency in evaluation metric scores when evaluating the diversity and quality of synthetic images across two biomedical imaging modalities: X-ray and Optical Coherence Tomography (OCT).
    \item Analyze the variability in evaluation metric scores when assessing the diversity and quality of synthetic images in both non-biomedical and biomedical imaging domains.
\end{itemize}

\section{Related Work}
\begin{table}[htp!]
\centering
\caption{\textbf{Evaluation of synthetic images using MS-SSIM, CD, and FID scores for assessing the quality and diversity of biomedical images. Sample size values refer to both real and synthetic images.}
}

\begin{tabular}{p{3.3cm}p{2.6cm}p{2.28cm}p{2cm}p{0.4cm}p{0.8cm}p{0.5cm}p{0.4cm}p{1cm}}

\toprule

\textbf{Reference} & \textbf{Med. Image}~\textsubscript{\textbf{Res.}} & \textbf{GANs} & \textbf{MS.} & \textbf{CD} & \textbf{FID} & \multicolumn{3}{|c}{\textbf{\textbf{Sample Size}}} \\
& & & & & & MS. & CD & FID \\
\midrule

\cite{saad2023self}
    & X-ray~\textsubscript{\textbf{128x128}}
    & MSG-SAGAN
    & 0.5, 0.47
    & N/A
    & 139.6
    & 3616
    & N/A
    & 3616
    \\
\cite{saad2022addressing}
    & X-ray~\textsubscript{\textbf{128x128}}
    & DCGAN
    & 0.5, 0.529
    & N/A
    & 0.687
    & 1340
    & N/A
    & 1340
    \\
\cite{qin2022learning}
    & X-ray~\textsubscript{\textbf{256x256}}
    & DCGAN
    & N/A
    & N/A
    & 293.26
    & N/A
    & N/A
    & N/S
    \\
\cite{segal2021evaluating}
    & X-ray~\textsubscript{\textbf{256x256}}
    & PGGAN
    & N/A
    & N/A
    & 8.02
    & N/A
    & N/A
    & 0.1 M
    \\
\cite{tajmirriahi2022dual}
    & OCT~\textsubscript{\textbf{128x128}}
    & DDFA-GAN
    & 0.17, 0.19
    & N/A
    & 51.30
    & N/S
    & N/A
    & N/S
    \\
\cite{he2020retinal}
    & OCT~\textsubscript{\textbf{224x224}}
    & LSGAN
    & N/A
    & N/A
    & N/A
    & N/A
    & N/A
    & N/A
    \\
\cite{segato2020data}
    & MRI~\textsubscript{\textbf{64x64}}
    & AEGAN
    & 0.99, 0.60
    & N/A
    & N/A
    & 2000
    & N/A
    & N/A
    \\
\bottomrule
\multicolumn{9}{l}{Med: Medical; N/S: Not Specified; MS: MS-SSIM (Real, Synthetic); M: Million} \\
\end{tabular}
\label{tab:GansMedicalImages}
\end{table}
Several studies on GANs have been proposed in the biomedical imaging domain, as indicated in Table \ref{tab:GansMedicalImages}. The majority of these studies have used FID scores to evaluate the quality of synthetic images. Few of which have used MS-SSIM to evaluate the intra-class diversity of synthetic images. Of those that quantify the FID, some studies such as \cite{qin2022learning} and \cite{tajmirriahi2022dual} do not specify the sample size used to evaluate it. Significant variance in FID scores is analyzed for X-ray images, see first four rows of Table \ref{tab:GansMedicalImages}. It is important to investigate the impact of the FID score, whether it depends on image size, image modality, GAN architecture, or sample size. Furthermore, none of the works reviewed considered whether different sample sizes of synthetic images could impact the evaluation of intra-class diversity and quality for these metrics. Similarly, in the non-biomedical imaging domains, studies reporting advances in GANs have failed to include details relating to the quantification of evaluation metrics, as indicated in Table. \ref{tab:GansNonMedicalImages}. 

In the literature, the research gap is highlighted in Table \ref{tab:GansMedicalImages} and Table \ref{tab:GansNonMedicalImages} to investigate the impact of sample size in evaluating the intra-class diversity and quality of synthetic images for biomedical and non-biomedical imaging domains. Prior studies also lack in demonstrating the variance in MS-SSIM, CD, and FID scores across biomedical-to-biomedical, and biomedical-to-non-biomedical imaging modalities.  

\begin{table}[htp!]
\centering
\caption{\textbf{Evaluation of synthetic images using MS-SSIM, CD, and FID scores for assessing the quality and diversity of synthetic images in the non-biomedical domain.}
}

\begin{tabular}{p{3.2cm}p{3cm}p{2.8cm}p{1.6cm}p{0.4cm}p{0.8cm}p{0.5cm}p{0.4cm}p{0.7cm}}

\toprule

\textbf{Reference} & \textbf{Img. Type}~\textsubscript{\textbf{Res.}} & \textbf{GANs} & \textbf{MS.} & \textbf{CD} & \textbf{FID} & \multicolumn{3}{|c}{\textbf{\textbf{Sample Size}}} \\
& & & & & & MS. & CD & FID \\
\midrule

\cite{allahyani2023divgan}
    & F-MNIST~\textsubscript{\textbf{28x28}}
    & DivGAN
    & N/A
    & N/A
    & 9.809
    & N/A
    & N/A
    & N/S
    \\
\cite{sanchez2023enhancing}
    & CelebA~\textsubscript{\textbf{128x128}}
    & REGAN
    & N/A
    & N/A
    & 36.57
    & N/A
    & N/A
    & 5K
    \\
\cite{lee2022generator}
    & LSUN-Church~\textsubscript{\textbf{256x256}}
    & StyleGAN2-GGDR
    & N/A
    & N/A
    & 3.15
    & N/A
    & N/A
    & 50K
    \\
\cite{yu2022hsgan}
    & CelebA~\textsubscript{\textbf{64x64}}
    & HSGAN
    & N/A
    & N/A
    & 17.49
    & N/A
    & N/A
    & N/S
    \\
\cite{zhang2021twgan}
    & F-MNIST~\textsubscript{\textbf{28x28}}
    & TWGAN
    & N/A
    & N/A
    & 10.3
    & N/A
    & N/A
    & N/S
    \\
\cite{zhao2021improved}
    & CelebA-HQ~\textsubscript{\textbf{128x128}}
    & BigGAN-ICR
    & N/A
    & N/A
    & 15.43
    & N/A
    & N/A
    & 3K
    \\
\cite{costa2020exploring}
    & CelebA~\textsubscript{\textbf{64x64}}
    & COEGAN-NSGC
    & N/A
    & N/A
    & 100
    & N/A
    & N/A
    & 1024
    \\
\cite{odena2017conditional}
    & Hot Dog~\textsubscript{\textbf{128x128}}
    & ACGAN
    & 0.11, 0.05
    & N/A
    & N/A
    & 200
    & N/A
    & N/A
    \\
\bottomrule
\multicolumn{9}{l}{N/S: Not Specified; MS: MS-SSIM (Real, Synthetic); N/A: Not Applied; K: Thousand} \\
\end{tabular}
\label{tab:GansNonMedicalImages}
\end{table}

\section{Methodology}


In this work, Chest X-ray and OCT images from the MedMNIST dataset \cite{yang2023medmnist} are used. The MedMNIST dataset contains several 28x28 resolution biomedical image datasets as a replica of the other benchmark datasets such as the F-MNIST dataset \cite{xiao2017fashion}, Pneumoniamnist, a binary class dataset containing normal and Pneumonia images, and OCTmnist, a multiclass dataset containing normal and three disease conditions are used, see Table \ref{tab:Distribution of X-ray dataset}. In the non-biomedical imaging domain, F-MNIST, a dataset containing different wearable products as indicated in Table \ref{tab:Distribution of F-mnist images} was used. All images were used with the original 28x28 resolution.     

\begin{table}[hbt!]
\centering
\caption{\textbf{Image distributions of Chest X-ray and Retinal OCT datasets.}}
\begin{tabular}{p{1cm}p{2.25cm}p{1cm}p{2cm}p{2.25cm}p{1cm}p{1cm}p{1cm}p{1cm}}
\toprule 
& \multicolumn{3}{c|}{\textbf{X-ray Images (Pneumoniamnist)}} & \multicolumn{5}{|c}{\textbf{Retinal OCT Images (OCTmnist)}} \\
& Total No. Img. & Normal & Pneumonia & Total No. Img. & Normal & Chor. & Diab. & Drusen \\
\midrule

Train & 4708 & 1214 & 3494 & 57884 & 27317 & 19867 & 6054 & 4646 \\
Valid & 524 & 135 & 389 & 10832 & 5114 & 3721 & 1135 & 862 \\
Test & 624 & 234 & 390 & 1000 & 250 & 250 & 250 & 250 \\
\bottomrule
\multicolumn{9}{l}{Img: Images; Chor: Choroidal Neovascularization; Diab: Diabetic Macular Edema} \\
\end{tabular}
\label{tab:Distribution of X-ray dataset}
\end{table}

\begin{table}[hbt!]
\centering
\caption{\textbf{Image distributions of Fashion MNIST dataset.}}
\begin{tabular}{p{0.8cm}p{2.4cm}p{0.8cm}p{0.6cm}p{0.6cm}p{0.6cm}p{1cm}p{0.8cm}p{0.8cm}p{1cm}p{0.8cm}p{0.6cm}}
\toprule
& \textbf{Total No. Img.} & \textbf{AB} & \textbf{Bg} & \textbf{Ct} & \textbf{Ds} & \textbf{Pr} & \textbf{St} & \textbf{Sl} & \textbf{Sr} & \textbf{T-st} & \textbf{Tr} \\
\midrule
Train & 60,000 & 6000 & 6000 & 6000 & 6000 & 6000 & 6000 & 6000 & 6000 & 6000 & 6000 \\
Test & 10,000 & 1000 & 1000 & 1000 & 1000 & 1000 & 1000 & 1000 & 1000 & 1000 & 1000 \\
\bottomrule
\multicolumn{12}{l}{Total No. Img: Total No. of Images per Class; AB: Ankle Boot; Bg: Bag; Ct: Coat; Ds: Dress; Pr: Pullover} \\
\multicolumn{12}{l}{St: Shirt; Sl: Sandal; Sn: Sneaker; T-st: T-shirt; Tr: Trouser} \\
\end{tabular}
\label{tab:Distribution of F-mnist images}
\end{table}

\subsection{DCGAN Architecture}

\begin{figure}[htp!]
    \centering
    \includegraphics[width=1\textwidth]{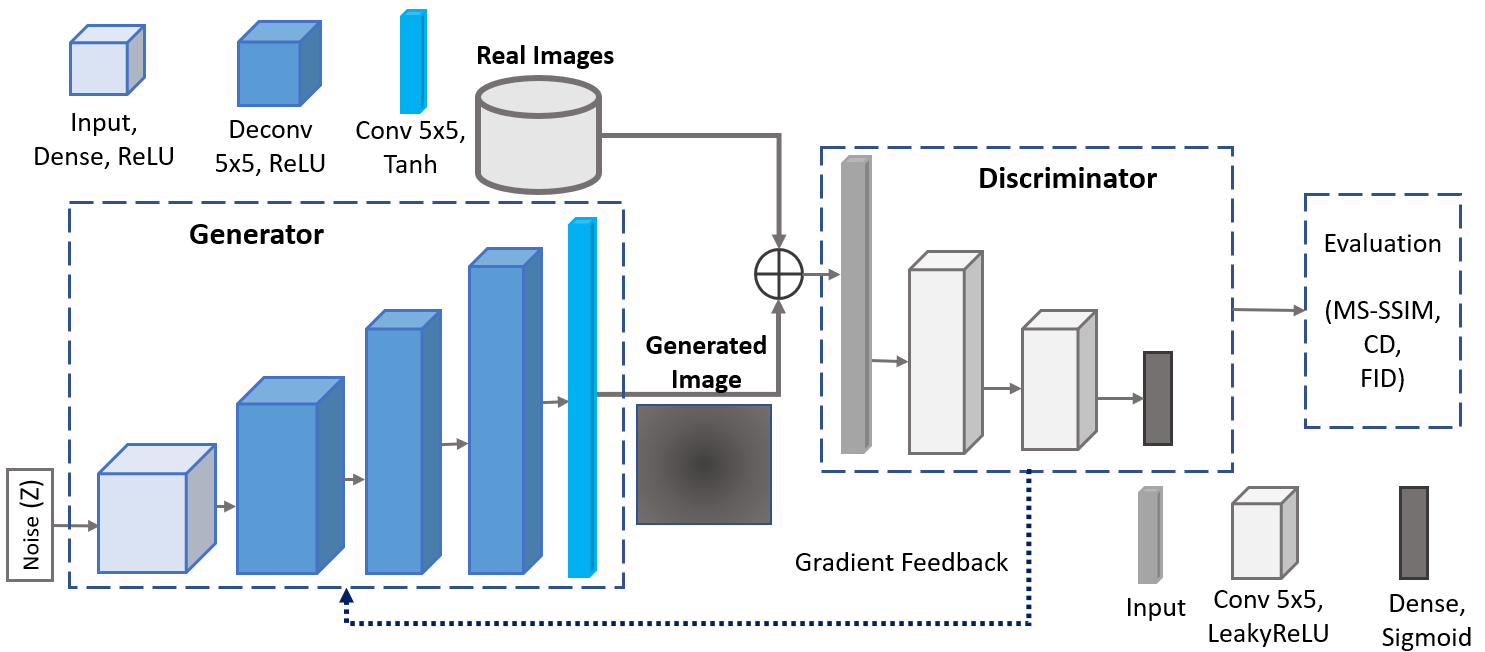}
    \caption{Architecture of DCGAN for synthesizing Chest X-ray, Retinal OCT, and Fashion MNIST images.}
    \label{Fig.dcgan_final}
\end{figure}

DCGAN is a baseline generative model, designed for generating realistic images. The DCGAN architecture implemented for the un-normalized X-ray images in \cite{saad2022addressing} was adapted and modified for 28x28 images used in this work. DCGAN architecture comprises two main components: the generator and the discriminator as detailed in Fig. \ref{Fig.dcgan_final}. The generator produces synthetic images using a latent input $z$ of 100 and passes them to the discriminator. The discriminator distinguishes synthetic images from real images and provides gradient feedback to the generator. The generator updates its learning based on the discriminator's gradient feedback to improve the generation of realistic and diverse images. DCGAN is trained for 500 epochs to converge the training to a balanced state with a batch size of 128. The binary-cross-entropy loss was used to evaluate the generator and discriminator performances. DCGAN is used for minority classes such as normal Chest X-ray images from the Pneumoniamnist, Drusen OCT images from the OCTmnist, and all F-MNIST images. DCGAN was trained for each class separately and generated synthetic images accordingly.

\subsection{Diversity of Synthetic Images}
MS-SSIM and CD are used to evaluate the intra-class diversity of synthetic images \cite{borji2019pros}, MS-SSIM computes the structural similarity at different levels, considering both local and global image features to evaluate the intra-class diversity of synthetic images. A higher MS-SSIM score of synthetic images than real images indicates that synthetic images lack diversity as compared to real images. Conversely, a lower MS-SSIM score of synthetic images indicates better diversity among the synthetic images as compared to real images. CD measures the distance between two feature vectors encoded by images to evaluate the intra-class diversity of synthetic images \cite{borji2019pros}. A higher CD among synthetic images indicates higher diversity compared to real images. Conversely, a lower CD among the synthetic images indicates limited diversity compared to real images.

\subsection{Quality of Synthetic Images}
FID measures the Wasserstein-2 distance between the two distributions (real and generated) using their mean vectors and covariance matrices to evaluate the quality of synthetic images compared to real images \cite{borji2022pros}. A lower FID score indicates that the synthetic images are closer in quality and distribution to the real images, indicating higher quality and better resemblance to the real images.

\subsection{Selecting Sample Size to Assess Intra-class Diversity and Quality of Synthetic Images}
\label{samplesize}
In this work, DCGAN has generated synthetic images equal to the number of real images (a 1:1 ratio). The impact of sample size quantifying the intra-class diversity and quality of synthetic images using different sample sizes such as 25\%, 50\%, 75\%, and 100\% (all 1:1 ratios) is assessed. 

\begin{figure}[hbt!]
  \centering 
\subfloat[Distribution of MS-SSIM scores across biomedical images.]{%
  \includegraphics[clip,width=0.5\textwidth,height=5cm]{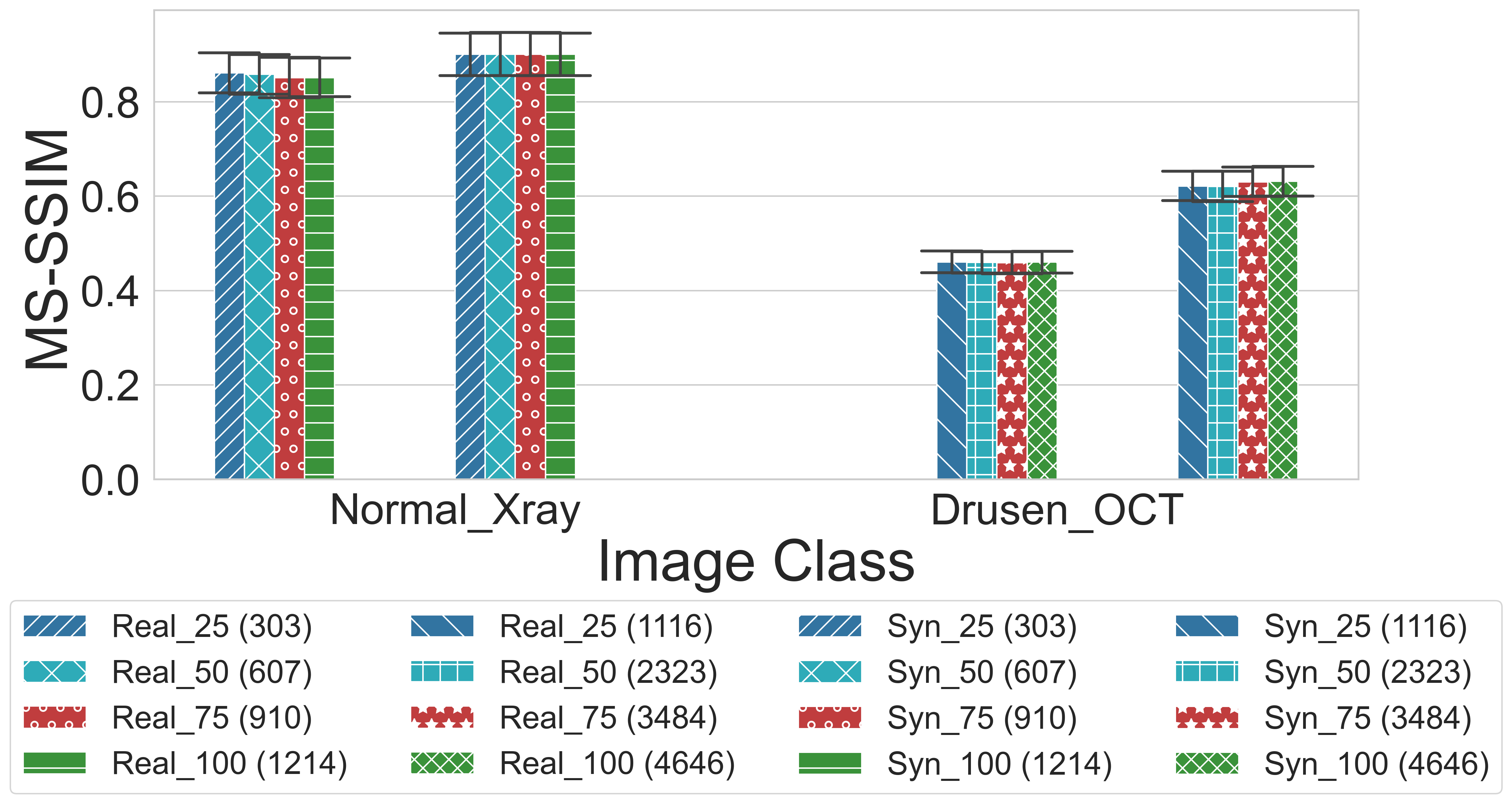}%
}
\subfloat[Distribution of CD scores across biomedical images.]{%
  \includegraphics[clip,width=0.5\textwidth,height=5cm]{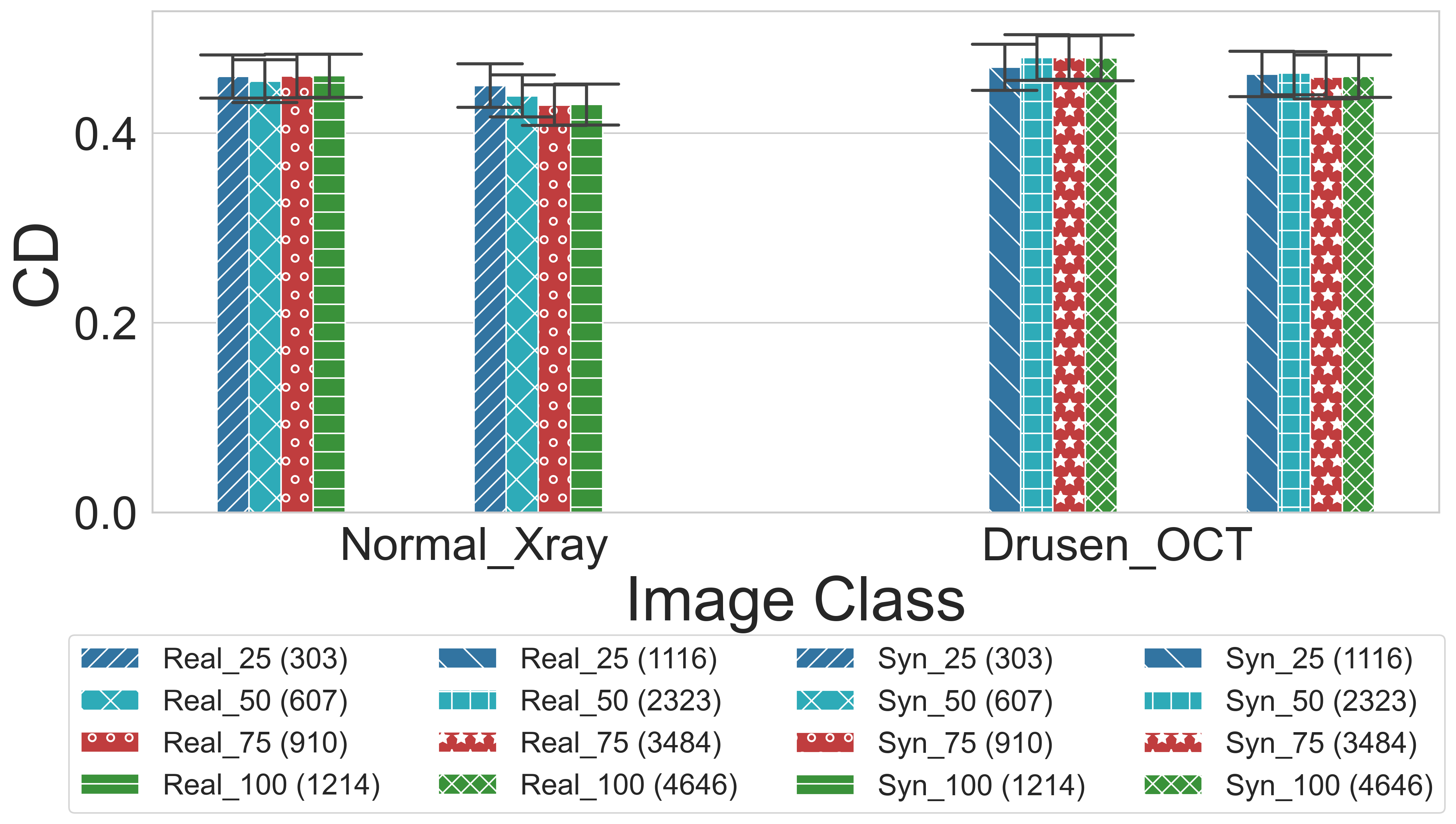}%
}

\caption{Intra-class diversity of biomedical images using different sample sizes.}
\label{MS-SSIM_CD_medical}
\end{figure}

\begin{figure}[hbt!]
  \centering   

\subfloat[Distribution of MS-SSIM scores across non-biomedical images.]{%
  \includegraphics[clip,width=1\textwidth,height=6cm]{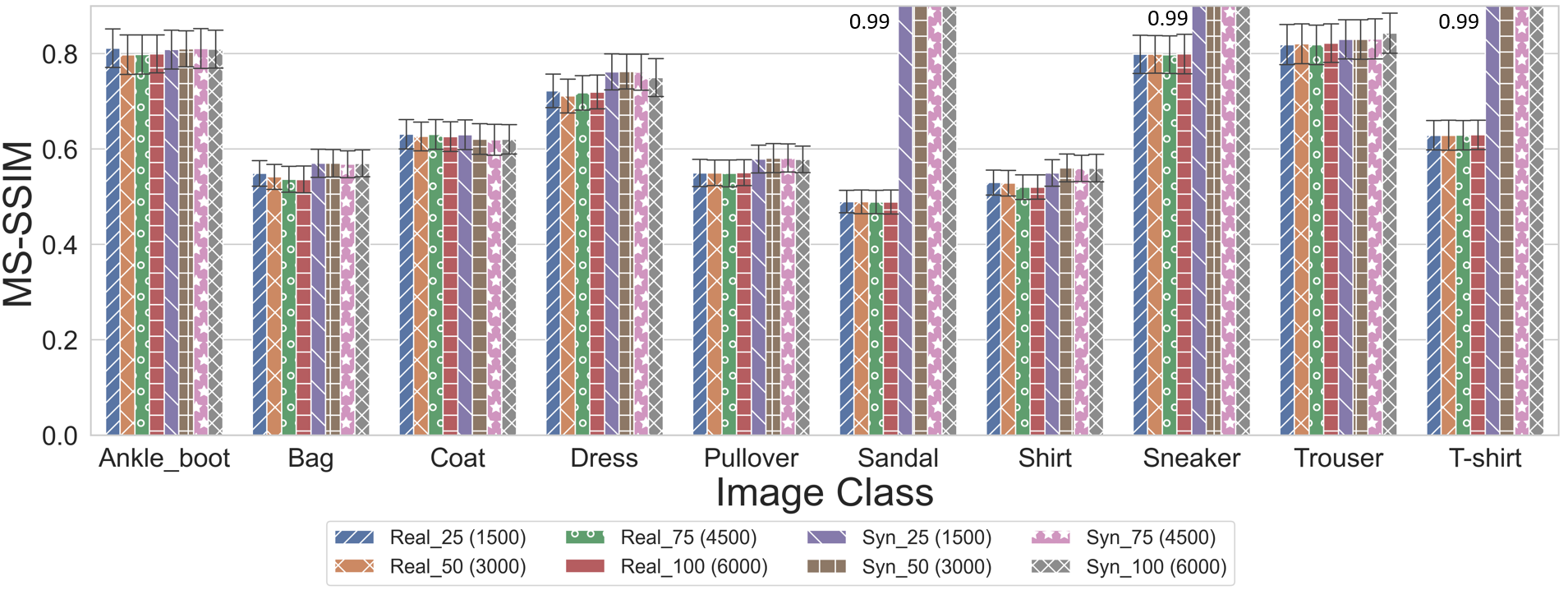}%
}

\subfloat[Distribution of CD scores across non-biomedical images.]{%
  \includegraphics[clip,width=1\textwidth,height=6cm]{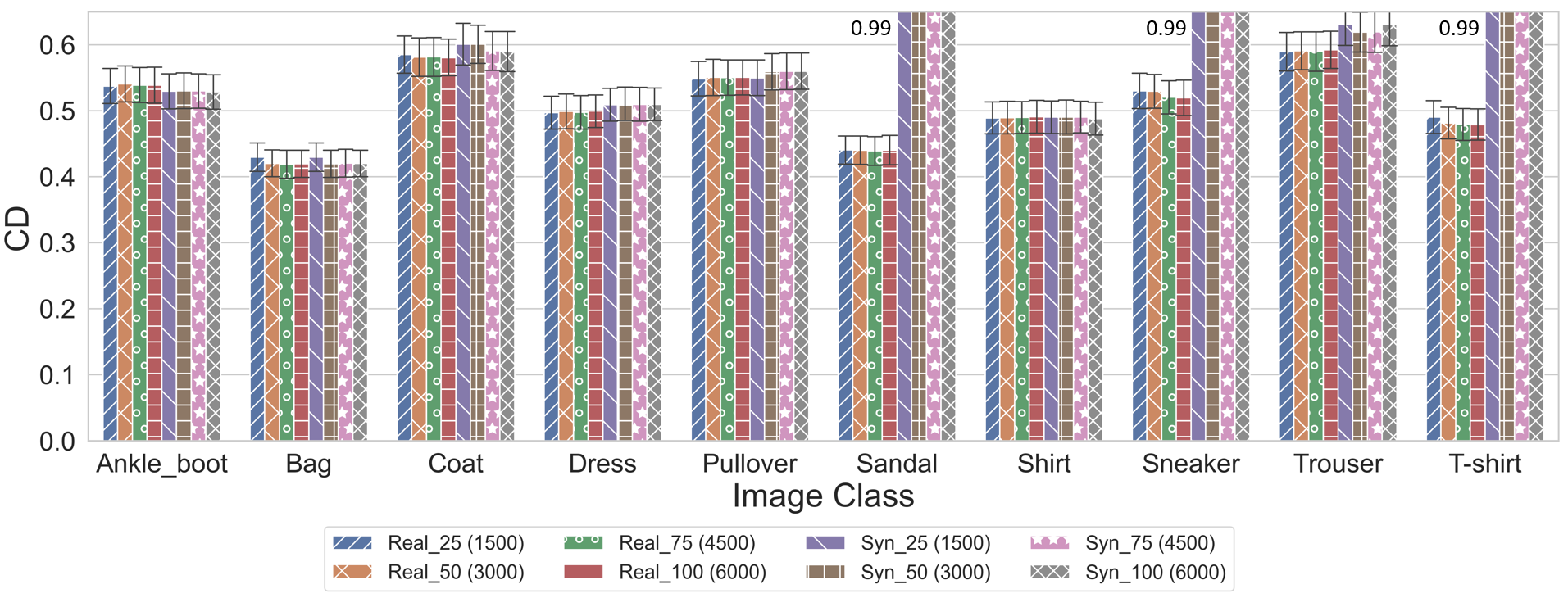}%
}
\caption{Intra-class diversity of non-biomedical images using different sample sizes.}
\label{MS-SSIM_CD_F-MNIST}
\end{figure}

\section{Results and Discussion}

\subsection{Variance in Diversity of Synthetic Images}
The intra-class diversity assessment of synthetic biomedical images using the MS-SSIM and CD scores is depicted in Fig. \ref{MS-SSIM_CD_medical}. The sample size has no significant impact on evaluating the intra-class diversity in X-ray and OCT images, as indicated by the uniform MS-SSIM and CD scores for different sample sizes. The MS-SSIM and CD scores vary across X-ray and OCT images because the distribution of features in images varies across X-ray and OCT modalities in real or synthetic images. 

The better MS-SSIM and CD scores of synthetic X-ray images over real images indicate better intra-class diversity of synthetic X-ray images. The poor MS-SSIM and CD scores of synthetic OCT images over real images indicate poor intra-class diversity of synthetic OCT images. MS-SSIM and CD scores of OCT images worsen over X-ray images because OCT images contain more diverse features than X-ray images, which are difficult to learn and train with the DCGAN. It also indicates that the architecture of GANs has a significant impact on the generation of synthetic images across different domains of biomedical imaging. 

For non-biomedical images, the MS-SSIM and CD analyses of real and synthetic Fashion MNIST images indicate significant variation across different classes, as depicted in Fig. \ref{MS-SSIM_CD_F-MNIST}. Each class has distinct images with unique features that impact the learning and training of DCGAN architecture. The assessment of different sample sizes for evaluating the MS-SSIM and CD scores indicates that there is no significant impact of sample size in evaluating these metric scores for intra-class diversity assessment. The analysis of these metric scores also indicates that the distribution of MS-SSIM and CD scores also varies across biomedical and non-biomedical images.

\subsection{Variance in Quality of Synthetic Images}
\begin{figure}[hbt!]
  \centering 
\subfloat[Distribution of FID scores across biomedical images.]{%
  \includegraphics[clip,width=0.45\textwidth,height=4cm]{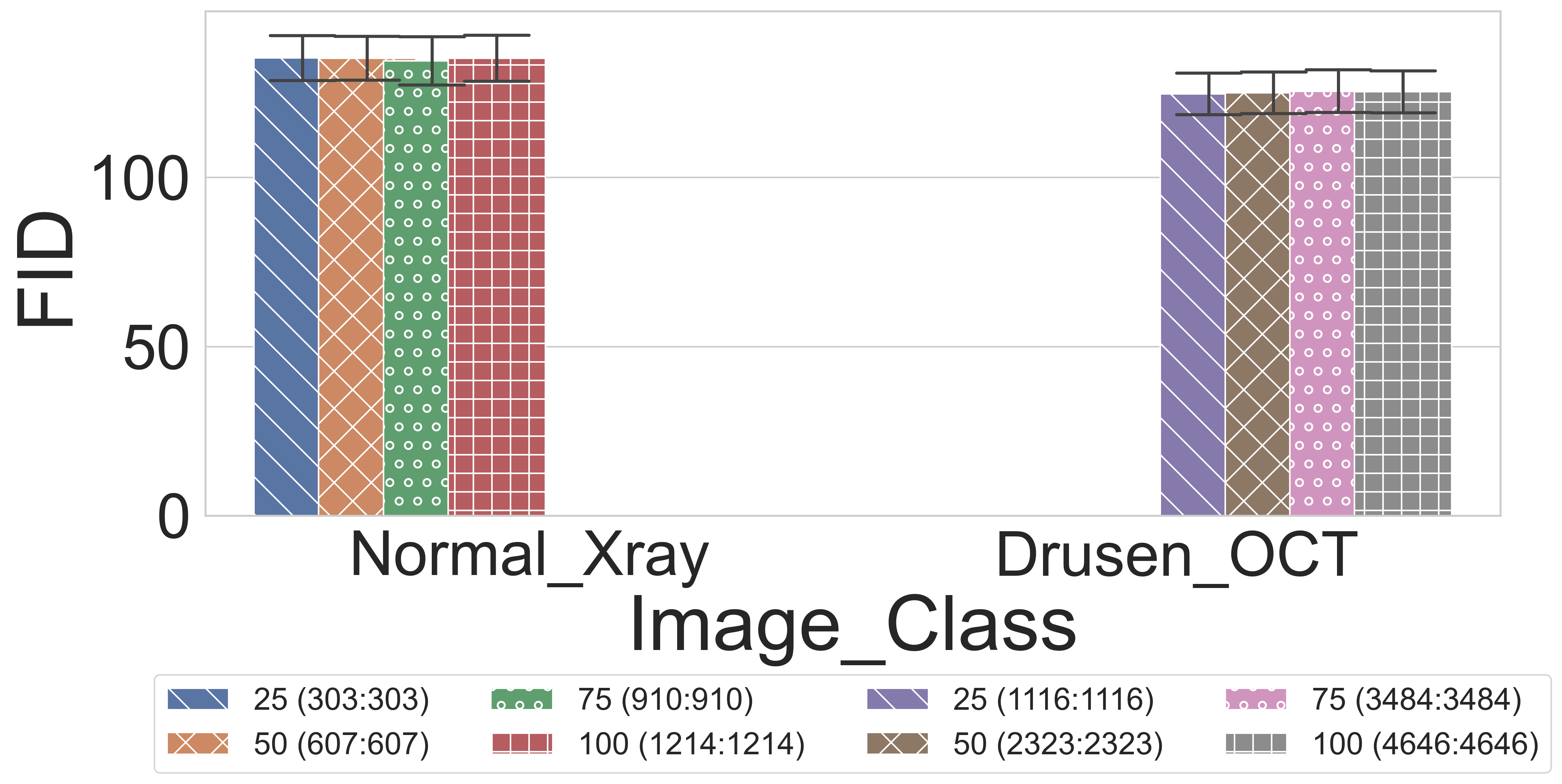}%
}
\subfloat[Distribution of FID scores across non-biomedical images.]{%
  \includegraphics[clip,width=0.55\textwidth,height=4cm]{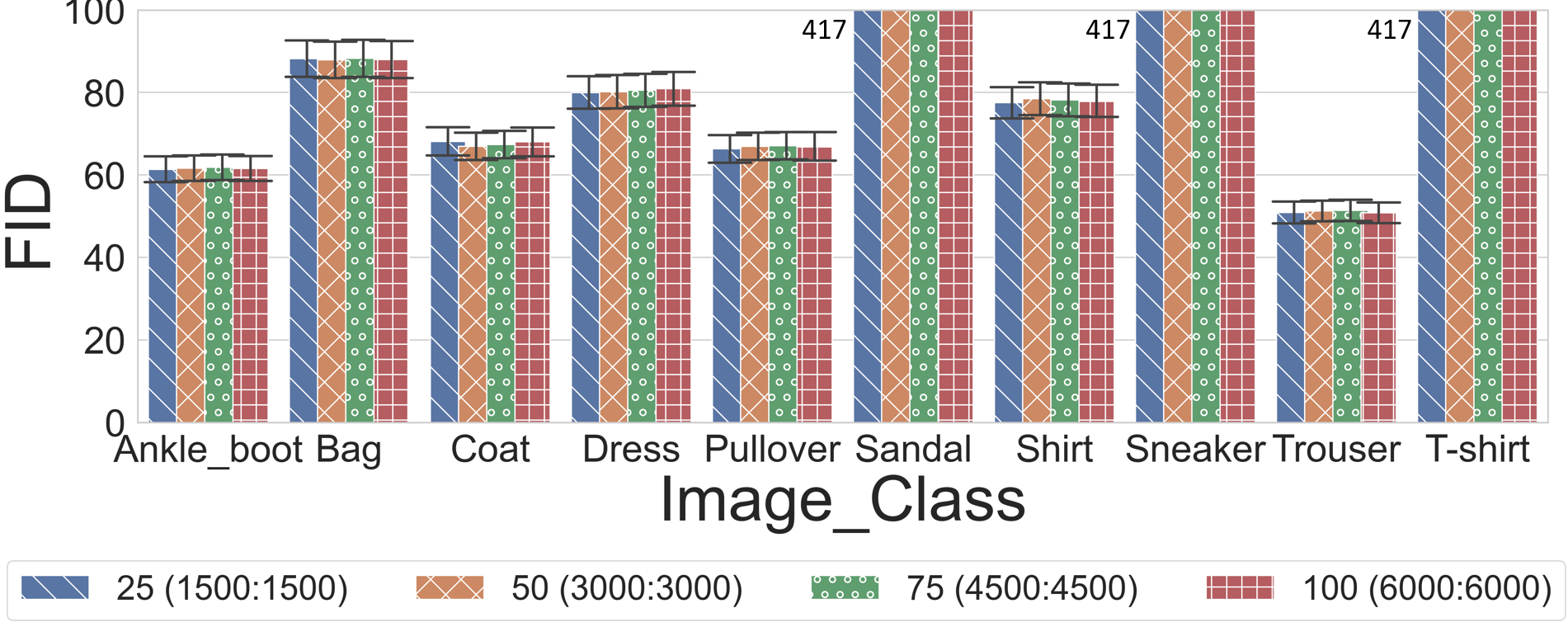}%
}
\caption{The quality of biomedical and non-biomedical images using different sample sizes.}
\label{FID_medical_F-MNIST}
\end{figure}

The assessment of the quality of synthetic images using the FID scores across biomedical images and non-biomedical images is indicated in Fig. \ref{FID_medical_F-MNIST}. The sample size of real and synthetic images has no significant impact on the FID scores in evaluating the quality of synthetic images across both biomedical and non-biomedical images as indicated by the uniform FID scores for different sample sizes. FID scores of normal X-ray, Drusen OCT, and Fashion MNIST images vary due to the distinct distribution of features in images of each biomedical and non-biomedical image class. FID has a different range of values across different biomedical and non-biomedical modalities. FID score can be higher or lower based on a specific biomedical image modality indicating the impact of salient features inherited in representative images.  

\section{Conclusion}

This work concludes that the intra-class diversity using the MS-SSIM and CD scores while the quality of synthetic images using FID scores vary across different biomedical imaging modalities due to the distribution of diverse and unique image features across different imaging modalities. 

Similarly, intra-class diversity and quality also vary across biomedical and non-biomedical imaging domains due to the distinct distribution of image features. Furthermore, the sample size has no significant impact on evaluating the intra-class diversity and quality of biomedical and non-biomedical images. One possible reason could be that synthetic images may exhibit limited diversity and quality as compared to real images. Therefore, varying the sample size does not impact the scores of evaluation metrics. Another reason could be a small image size, whereby varying the sample size does not impact the metric scores.  

The minimum sample size should be a few hundred images of the real and synthetic images to measure the MS-SSIM, CD, and FID metrics. This research work also shows that the generation of synthetic images using DCGAN is dependent upon the nature of images to produce diversified and high-quality images across different imaging modalities in biomedical and non-biomedical domains.



\end{document}